\documentstyle[jkas]{article}
\newcommand{\beqe}{\begin{equation}} \newcommand{\eeqe}{\end{equation}}
\newcommand{\p}[1]{(\ref{#1})}
\beginpage{1}
\endpage{11}

\year{2010} \volume{39}

\runningauthor {A.A. ISAYEV \& J. YANG} \year{2010} \volume{39}
\beginpage{1}\endpage{8}
\runningtitle{FINITE TEMPERATURE EFFECTS ON SPIN POLARIZATION }
\begin{document}
\title{FINITE TEMPERATURE EFFECTS ON SPIN POLARIZATION \\OF NEUTRON MATTER IN A STRONG
MAGNETIC FIELD}
\author{Alexander A. Isayev$^{1}$, and Jongmann Yang$^2$}
\address{$^1$ Kharkov Institute of
Physics and Technology, Academicheskaya Street 1,
 Kharkov, 61108, Ukraine\\
 {\it e-mail}: isayev@kipt.kharkov.ua}
\address{$^2$ Department  of Physics and the Institute for the Early Universe,
 \\
Ewha Womans University, Seoul 120-750, Korea\\ {\it e-mail}: jyang@ewha.ac.kr}

\address{\normalsize{\it (Received August XXX, 2010; Accepted XXX)}}
%\offprints{J. Yang}
%--------------------------------------------------------------------
\abstract{Magnetars are neutron stars possessing magnetic field of
about $10^{14}$-$10^{15}$~G at the surface. Thermodynamic properties
of neutron star matter approximated by pure neutron matter are
considered at finite temperature in strong magnetic fields up to
$10^{18}$~G which could be relevant for the interior regions of
magnetars. In the model with the Skyrme effective interaction, it is
shown that a thermodynamically stable branch of solutions for the
spin polarization parameter corresponds to the case  when the
majority of neutron spins are oriented opposite to the direction of
the magnetic field (negative spin polarization). Besides, the
self-consistent equations, beginning from some threshold density,
have also two other branches of solutions corresponding to positive
spin polarization. The influence of finite temperatures on spin
polarization remains moderate in the Skyrme model up to temperatures
relevant for protoneutron stars. In particular, the scenario with
the metastable state characterized by positive spin polarization,
considered at zero temperature in Phys. Rev. C {\bf 80}, 065801
(2009), is preserved at finite temperatures as well. It is shown
that above certain density the entropy for various branches of spin
polarization in neutron matter with the Skyrme interaction in a
strong magnetic field demonstrates the unusual behavior being larger
than that of the nonpolarized state. By providing the corresponding
low-temperature analysis, it is clarified  that this unexpected
behavior should be addressed to the dependence of the entropy of a
spin polarized state on  the effective masses of neutrons with spin
up and spin down, and to a certain constraint on them which is
violated in the respective density range. }

  \maketitle

%--------------------------------------------------------------------
\section{INTRODUCTION}
Magnetars are strongly magnetized neutron stars (\cite{DT}) with
the magnetic field strength at the surface of about
$10^{14}$-$10^{15}$~G (\cite{TD,IShS}). Such huge magnetic fields
can be inferred from observations of magnetar periods and
spin-down rates, or from hydrogen spectral lines. Among possible
classes of various neutron stars, soft gamma-ray repeaters and
anomalous X-ray pulsars are believed to be most probable
candidates for these ultrastrong magnetized astrophysical bodies
(\cite{WT}). Magnetars are relatively frequent objects in the
Universe and comprise about $10\%$ of the whole population of
neutron stars (\cite{K}). In the interior of a magnetar the
magnetic field strength may be even larger, reaching values of
about $10^{18}$~G (\cite{CBP,BPL}). Therefore, magnetars provide a
unique playground for studying  properties of neutron star matter
under extreme conditions of density and magnetic field strength
(\cite{CBP,BPL,CPL,PG,IY09}), which are inaccessible in the
terrestrial laboratories.

In the recent study by Perez-Garcia~(2008), neutron star matter was
approximated by  pure neutron matter in a model with the effective
nuclear  forces. It has been shown that the behavior of spin
polarization of neutron matter in the high density region in a
strong magnetic field crucially depends on whether neutron matter
develops a spontaneous spin polarization (in the absence of a
magnetic field) at  several times  nuclear matter saturation
density, or the appearance of a spontaneous polarization is not
allowed  at the relevant densities (or delayed to much higher
densities). The first case  is usual for the Skyrme
forces~(\cite{R,S,O,VNB,RPLP,ALP,MNQN,TT,BPM,KW94,I,IY04a,RPV,I06}),
while the second one is characteristic  for the realistic
nucleon-nucleon (NN) interaction~(\cite{PGS,BK,H,VPR,FSS,KS,BB}). In
the former case, a ferromagnetic transition to a totally spin
polarized state occurs while in the latter case a ferromagnetic
transition is excluded at all relevant densities and the spin
polarization remains quite low even in the high density region. If a
spontaneous ferromagnetic transition is allowed,   it was shown in
the subsequent model consideration with the Skyrme effective forces
(~\cite{IY09}) that the self-consistent equations for the spin
polarization parameter at nonzero magnetic field have not only
solutions corresponding to negative spin polarization (with the
majority of neutron spins oriented opposite to the direction of the
magnetic field) but, because of the strong spin-dependent medium
correlations in the high-density region, also the solutions with
positive spin polarization. In the last case, the formation of a
metastable state with the majority of neutron spins oriented along
the magnetic field is possible in the high-density interior of a
neutron star.

In this study, we extend our previous consideration of neutron
matter with Skyrme forces in a strong magnetic field, given at
zero temperature~(\cite{IY09}), to finite temperatures up to a few
tens of MeV being relevant for protoneutron stars. As a framework
for consideration, we use a Fermi liquid approach for the
description of nuclear matter~(\cite{AKPY,AIP,IY3}). One of the
goals of the research is to study the impact of finite
temperatures on spin polarized states in neutron matter in a
strong magnetic field, and, in particular, to clarify whether the
formation of a metastable state with positive spin polarization is
possible in the high-density region of neutron matter. Besides, we
provide a fully self-consistent finite temperature calculation of
the basic thermodynamic functions of spin polarized neutron matter
in a strong magnetic field. It will be shown that beginning from
some density the entropy for various branches of spin polarization
in neutron matter in a strong magnetic field demonstrates the
unusual behavior being larger than that of nonpolarized neutron
matter. We relate this unexpected result to the dependence of the
entropy on the effective masses of neutrons with spin up and spin
down, and to the violation of a certain constraint on them in the
corresponding density range.

Note that we consider  thermodynamic properties of spin polarized
states in neutron  matter in a strong magnetic field up to the
high density region relevant for astrophysics. Nevertheless, we
take into account the nucleon degrees of freedom only, although
other degrees of freedom, such as pions, hyperons, kaons, or
quarks could be important at such high densities.

\section{BASIC EQUATIONS}
 Here we present only the basic formulae necessary for further
 calculations, although more details concerning a Fermi-liquid
 approach to neutron matter in a strong magnetic field
 can be found in our earlier work (\cite{IY09}).
 The normal (nonsuperfluid) states of neutron matter are described
  by the normal distribution function of neutrons $f_{\kappa_1\kappa_2}=\mbox{Tr}\,\varrho
  a^+_{\kappa_2}a_{\kappa_1}$, where
$\kappa\equiv({\bf{p}},\sigma)$, ${\bf p}$ is momentum, $\sigma$
is the projection of spin on the third axis, and $\varrho$ is the
density matrix of the system. Further it will be assumed that the
third axis is directed along the external magnetic field $\bf{H}$.
Given the possibility for alignment of neutron spins along or
opposite to the magnetic field $\bf H$, the normal distribution
function of neutrons and single particle energy can be expanded in
the Pauli matrices $\sigma_i$ in spin
space%~\cite{AIP}
\begin{eqnarray} f({\bf p})&=& f_{0}({\bf
p})\sigma_0+f_{3}({\bf p})\sigma_3,\label{7.2}\\
\varepsilon({\bf p})&=& \varepsilon_{0}({\bf
p})\sigma_0+\varepsilon_{3}({\bf p})\sigma_3.
 \nonumber
\end{eqnarray}

Expressions for  the distribution functions $f_{0},f_{3}$
 in
terms of the quantities $\varepsilon$ read~(\cite{I,IY04a})

 \begin{eqnarray}
f_{0}&=&\frac{1}{2}\{n(\omega_{+})+n(\omega_{-}) \},\label{2.4}
 \\
f_{3}&=&\frac{1}{2}\{n(\omega_{+})-n(\omega_{-})\}.\nonumber
 \end{eqnarray} Here $n(\omega)=\{\exp(Y_0\omega)+1\}^{-1}$ and
 \begin{eqnarray}
\omega_{\pm}&=&\xi_{0}\pm\xi_{3},\label{omega}\\
\xi_{0}&=&\varepsilon_{0}-\mu_{0},\;
\xi_{3}=\varepsilon_{3},\nonumber\end{eqnarray} $\mu_0$ being the
chemical potential of neutrons. The branches  $\omega_{\pm}$ of
the quasiparticle spectrum correspond to neutrons with spin up and
spin down.

The distribution functions $f$ should satisfy the norma\-lization
conditions
\begin{eqnarray} \frac{2}{\cal
V}\sum_{\bf p}f_{0}({\bf p})&=&\varrho,\label{3.1}\\
\frac{2}{\cal V}\sum_{\bf p}f_{3}({\bf
p})&=&\varrho_\uparrow-\varrho_\downarrow\equiv\Delta\varrho.\label{3.2}
 \end{eqnarray}
 Here $\varrho=\varrho_{\uparrow}+\varrho_{\downarrow}$ is the total density of
 neutron matter, $\varrho_{\uparrow}$ and $\varrho_{\downarrow}$  are the neutron number densities
 with spin up and spin down,
 respectively. The
quantity $\Delta\varrho$  may be regarded as the neutron spin
order parameter. It determines the magnetization of the system
$M=\mu_n \Delta\varrho$, $\mu_n$ being the neutron magnetic
moment. The magnetization may contribute to the internal magnetic
field $B=H+4\pi M$. However, as we discussed
earlier~(\cite{IY09}), and, analogously to the previous works
(Perez-Garcia 2008, Broderick, Prakash \& Lattimer 2000),
 we will assume %, analogously to
%Refs.~\cite{PG,BPL},
that the contribution of the magnetization
 to the magnetic field
$B$ remains small for all relevant densities and magnetic field
strengths, and, hence, $ B\approx H$.

The self-consistent equations for the components of the
single-particle energy have the form~(\cite{IY09})
\begin{eqnarray}\xi_{0}({\bf
p})&=&\underline\varepsilon_{\,0}({\bf
p})+\tilde\varepsilon_{0}({\bf p})-\mu_0,\label{14.1}\\
\xi_{3}({\bf p})&=&-\mu_nH+\tilde\varepsilon_{3}({\bf
p}).\label{14.2}
\end{eqnarray}
Here  $\underline{\varepsilon}_{\,0}({\bf p})=\frac{{\bf
p}^{\,2}}{2m_{0}}$ is the free single particle spectrum, $m_0$ is
the bare mass of a neutron, and
$\tilde\varepsilon_{0},\tilde\varepsilon_{3}$ are the Fermi liquid
(FL) corrections to the free single particle spectrum, related to
the normal FL amplitudes $U_0^n({\bf k}), U_1^n({\bf k})$ by
formulas
\begin{eqnarray}\tilde\varepsilon_{0}({\bf p})&=&\frac{1}{2\cal
V}\sum_{\bf q}U_0^n({\bf k})f_{0}({\bf
q}),\;{\bf k}=\frac{{\bf p}-{\bf q}}{2}, \label{flenergies}\\
\tilde\varepsilon_{3}({\bf p})&=&\frac{1}{2\cal V}\sum_{\bf
q}U_1^n({\bf k})f_{3}({\bf q}).
\end{eqnarray}

To obtain
 numerical results, we  utilize the  effective Skyrme interaction~(\cite{VB}).
 Expressions for the normal FL amplitudes in terms of the Skyrme
  force parameters were written by~\cite{AIP,IY3}.
Thus, using expressions~\p{2.4} for the distribution functions
$f$, we obtain the self-consistent equations~\p{14.1}, \p{14.2}
for the components of the single-particle energy $\xi_{0}({\bf
p})$ and $\xi_{3}({\bf p})$, which should be solved jointly with
the normalization conditions~\p{3.1}, \p{3.2}.

 Note that  spin ordering in
neutron matter can be characterized by the  neutron spin
polarization parameter $$ %\beqe
\Pi=\frac{\varrho_{\uparrow}-\varrho_{\downarrow}}{\varrho}\equiv\frac{\Delta\varrho}{\varrho}.
$$ %\end{equation}
The  number densities of neutrons with spin up and spin down are
related to the spin polarization parameter $\Pi$ by formulas
\begin{equation}
\varrho_{\uparrow}=\frac{\varrho}{2}(1+\Pi),\;\varrho_{\downarrow}=\frac{\varrho}{2}(1-\Pi).
\label{rho}\end{equation}

To examine the thermodynamic stability of different solutions of
the self-consistent equations, it is necessary to compare the
corresponding free energies $F=E-TS$, where the energy functional
$E$ is characterized by two FL amplitudes
$U_0^n,U_1^n$~(\cite{IY3}) and the entropy reads
\begin{eqnarray} S&=&-\sum_{\bf
p}\,\sum_{\sigma=+,\,-}\{n(\omega_{\sigma})\ln
n(\omega_{\sigma})\label{entr}\\ &&+\bar n(\omega_{\sigma})\ln
\bar n(\omega_{\sigma})\}, \;\bar n(\omega)=1-n(\omega).\nonumber
\end{eqnarray}

\section{SOLUTIONS OF SELF-CONSISTENT\\ EQUATIONS AT FINITE $T$. THERMODYNAMIC FUNCTIONS}

The self-consistent equations were analyzed at zero temperature
by~\cite{IY09} for the magnetic field strengths up to
$H_{max}\sim10^{18}$~G, allowed by a scalar virial
theorem~(\cite{LS}), in the model consideration with SLy4 and SLy7
Skyrme effective forces~(\cite{CBH}). These Skyrme
parametrizations were constrained originally to reproduce the
results of microscopic neutron matter calculations
(pressure-versus-density curve) and give neutron star models in a
broad agreement with the observables such as the minimum rotation
period, gravitational mass-radius relation, the binding energy,
released in supernova collapse, etc.~(\cite{RMK}). It was shown
that a thermodynamically stable branch of solutions for the
spin-polarization parameter as a function of density corresponds
to  negative spin polarization when the majority of neutron spins
are oriented opposite to the direction of the magnetic field.
Besides,  beginning from some threshold density dependent on the
magnetic field strength,  the state with  positive spin
polarization can be realized as a metastable state in the
high-density region of neutron matter in the model with the Skyrme
effective forces.

Here we  study the impact of finite temperatures on spin
polarization of neutron matter
 in a strong magnetic field. In
particular, we are interested to learn whether the metastable
state with  positive spin polarization will survive at
temperatures about several tens of MeV relevant for protoneutron
stars. To that end, we directly find solutions of the
self-consistent equations at nonzero temperature.  Because the
results of calculations with SLy4 and SLy7 Skyrme forces are very
close, here we present the obtained dependences only for the SLy7
Skyrme interaction.

\begin{figure}[t]
\epsfxsize=\linewidth \epsfbox{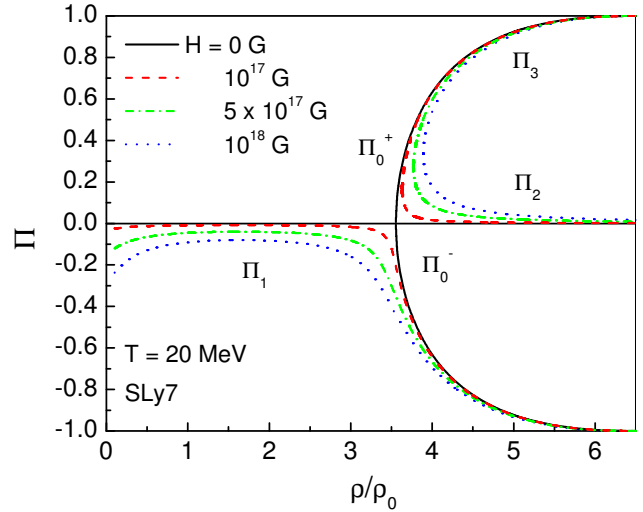} \caption{(Color online)
Neutron spin polarization parameter as a function of density at
$T=20$~MeV and different magnetic field strengths for SLy7
interaction. The branches of spontaneous polarization
$\Pi_0^-,\Pi_0^+$ are shown by solid
curves.}\label{fig1}\vspace{-0ex}
\end{figure}

Fig.~\ref{fig1} shows the neutron spin polarization parameter  as
a function of density, normalized to the nuclear saturation
density $\varrho_0$ (for SLy7 force,
$\varrho_0=0.158\,\mathrm{fm}^{-3}$), for a set of fixed values of
the magnetic field strength at the temperature $T=20$~MeV. The
branches of spontaneous polarization $\Pi_0^-,\Pi_0^+$,
corresponding to the vanishing magnetic field, are shown by solid
curves. As one can see, the obtained dependences qualitatively are
similar to those obtained at zero temperature~\cite{IY4}. The
branches of spontaneous polarization are modified differently by
the magnetic field: The branch $\Pi_0^-(\varrho)$ turns to the
branch $\Pi_1(\varrho)$ with negative spin polarization while the
branch $\Pi_0^+(\varrho)$ splits into two branches,
$\Pi_2(\varrho)$ and $\Pi_3(\varrho)$, corresponding to positive
spin polarization. For the lower branch $\Pi_1(\varrho)$, there
are three characteristic density domains. At low densities
$\varrho\lesssim 0.5\varrho_0$, the magnitude
 of the spin polarization parameter increases with
decreasing density. At intermediate densities
$0.5\varrho_0\lesssim\varrho\lesssim3\varrho_0$, there is a
plateau in the $\Pi_1(\varrho)$  dependence, whose characteristic
value depends on $H$ (at the given temperature). At densities
$\varrho\gtrsim3\varrho_0$,  the absolute value of the spin
polarization parameter increases with density and tends to unity.
Note that in the
 low-density domain the possibility of the appearance of a "nuclear
magnetic pasta" and its impact  on the neutrino opacities in the
protoneutron star early cooling stage should be explored in a more
elaborated  analysis as  discussed in detail by~Perez-Garcia
(2008).

The upper branches $\Pi_2(\varrho)$ and $\Pi_3(\varrho)$,
corresponding to positive spin polarization, appear stepwise at
the same threshold density $\varrho_{\rm th}$ dependent on the
magnetic field (at the given temperature) and being larger than
the critical density of spontaneous spin instability in neutron
matter.  For the branch $\Pi_2(\varrho)$, the spin polarization
parameter decreases with density and tends to zero while for the
branch $\Pi_3(\varrho)$ it increases with density and approaches
unity. Because of the negative value of the neutron magnetic
moment, the magnetic field  tends to orient the neutron spins
opposite to the magnetic field direction. As a result, the spin
polarization parameter for the branches $\Pi_2(\varrho)$,
$\Pi_3(\varrho)$ with positive spin polarization is smaller than
that for the branch of spontaneous polarization $\Pi_0^+$, and,
vice versa, the magnitude of the spin polarization parameter for
the branch $\Pi_1(\varrho)$ with negative spin polarization is
larger than the corresponding value for the branch of spontaneous
polarization $\Pi_0^-$.

\begin{figure}[t]
\epsfxsize=\linewidth \epsfbox{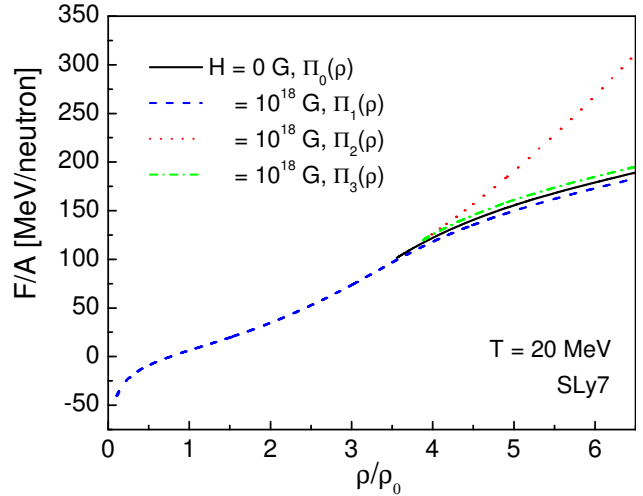} \caption{(Color online)
Free energy  per neutron  as a function of density at $T=20$~MeV
for different branches $\Pi_1(\varrho)$-$\Pi_3(\varrho)$ of
solutions of the self-consistent equations at $H=10^{18}$~G for
SLy7 interaction, including a spontaneously polarized
state.}\label{fig2}\vspace{-0ex}
\end{figure}

Thus, at densities larger than $\varrho_{\rm th}$, we have three
branches of solutions: one of them, $\Pi_1(\varrho)$,  with
negative spin polarization and two others, $\Pi_2(\varrho)$ and
$\Pi_3(\varrho)$, with  positive polarization. In order to clarify
which branch is thermodynamically preferable we should compare the
corresponding free energies. Fig.~\ref{fig2} shows the free energy
per neutron as a function of density at $T=20$~MeV and
$H=10^{18}$~G for these three branches, compared with the free
energy per neutron for a spontaneously polarized state [the
branches $\Pi_0^\pm(\varrho)$]. It is seen that the state with the
majority of neutron spins  oriented opposite to the direction of
the magnetic field [the branch $\Pi_1(\varrho)$] has a lowest free
energy. This result is intuitively clear, since magnetic field
tends to direct the neutron spins opposite to $\bf{H}$, as
mentioned earlier. However, the state, described by the branch
$\Pi_3(\varrho)$ with  positive spin polarization, has the free
energy very close to that of the thermodynamically stable state.
This means that despite the presence of a strong magnetic field
$H\sim 10^{18}$~G, the state with the majority of neutron spins
directed  along the magnetic field can be realized as a metastable
state in the dense core of a neutron star in the model
consideration with the Skyrme effective interaction. In this
scenario, because such states exist only at densities
$\varrho\geq\varrho_{\rm th}$,
 under decreasing density (going from the
interior to the outer regions of a magnetar) a metastable state
with  positive spin polarization  at the threshold density
$\varrho_{\rm th}$ changes to a thermodynamically stable state
with negative spin polarization.

Note at this point that the possibility of the appearance of a
metastable state with the majority of neutron spins oriented along
the magnetic field was missed in the finite temperature
calculations by Perez-Garcia (2008) because only one branch of
solutions for the spin polarization parameter, corresponding to
the thermodynamically stable state,  was found there for neutron
matter in a strong magnetic field with the same Skyrme
interaction.

\begin{figure}[t]
\epsfxsize=\linewidth \epsfbox{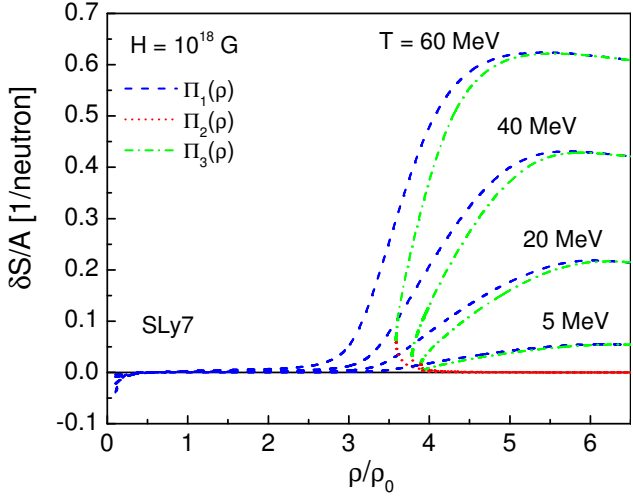} \caption{(Color online)
 The entropy per neutron
measured from its value in the nonpolarized state for the
$\Pi_1$--$\Pi_3$ branches of spin polarization as a function of
density at $H=10^{18}$~G and different temperatures for the SLy7
interaction.}\label{fig3}
\end{figure}

An unexpected moment appears if we consider separately the entropy
for various branches of spin polarization. Fig.~\ref{fig3} shows
the difference between the entropy per neutron for the branches
$\Pi_1$--$\Pi_3$ and that of the nonpolarized state (with $\Pi=0$
at $H=0$) as a function of density at the fixed magnetic field
strength $H=10^{18}$~G and different temperatures. Contrarily to
the intuitively expected behavior, the entropy for the branch
$\Pi_1$, beginning from a certain density  weakly dependent on
temperature (at the given magnetic field strength), is larger than
the entropy of the nonpolarized state. Besides,  the entropy for
the branches $\Pi_2$ and $\Pi_3$ is larger than that for the
nonpolarized state for all relevant densities. It looks like the
spin polarized states described by the $\Pi_1$--$\Pi_3$ branches
of spin polarization are less ordered than the nonpolarized state
for the corresponding densities.

In order to understand qualitatively such an expected behavior,
let us consider the low-temperature expansion for the entropy in
terms of the effective masses of spin-up and spin-down neutrons.
The density of entropy~\p{entr} can be written in the form

\begin{eqnarray}s&=&\frac{1}{\pi^2\hbar^3}
\sum_{\sigma=+,\,-}\sqrt{\frac{m_\sigma^3T^3}{2}}\label{denentr}\\
&&\times\biggl\{\frac{5}{3} J_\frac{3}{2}(\eta_\sigma)-\eta_\sigma
J_{\frac{1}{2}}(\eta_\sigma)\biggr\},\;
\eta_\sigma\equiv\frac{\mu_\sigma}{T},\nonumber\end{eqnarray}
where $$ J_\nu(\eta)=\int_0^\infty\frac{x^\nu}{e^{x-\eta}+1}dx$$
is Fermi-Dirac integral of the order $\nu$, $\mu_\sigma$ is the
effective chemical potential of neutrons with spin up ($\sigma=+$)
and spin down($\sigma=-$), whose explicit expression was written
by Isayev \& Yang (2009). In the low-temperature limit,
$\eta_\sigma\gg1$, and, providing the corresponding expansion  of
Fermi-Dirac integrals in Eq.~\p{denentr}, one gets
\begin{equation}s=\sum_{\sigma=+,\,-}s_\sigma,\quad
s_\sigma=\frac{\pi^2}{2\varepsilon_{F\sigma}}T,\label{lowlim}\end{equation}
where $\varepsilon_\sigma=\frac{\hbar^2k_{F\sigma}^2}{2m_\sigma}$ is
the Fermi energy of neutrons with spin up and spin down,
$k_\sigma=(6\pi^2\varrho_\sigma)^{1/3}$ being  the respective Fermi
momentum. The low-temperature expansion~\p{lowlim} is valid till
$T/\varepsilon_{F\sigma}\ll 1$. Then, requiring for the difference
between the entropies of spin polarized and nonpolarized states to
be negative, one gets the constraint on the effective masses
$m_{n\uparrow}$ and $m_{n\downarrow}$ of neutrons with spin up and
spin down in a spin polarized state:
\begin{equation}D\equiv
\frac{m_{n\uparrow}}{m_n}(1+\Pi)^\frac{1}{3}+
\frac{m_{n\downarrow}}{m_n}(1-\Pi)^\frac{1}{3}-2<0,\label{lowtemD}\end{equation}
Here $m_n$ is the effective mass of a neutron in nonpolarized
neutron matter (\cite{IY04a}), \begin{equation}
\frac{\hbar^2}{2m_{n}}=\frac{\hbar^2}{2m_0}+\frac{\varrho}{8}
[t_1(1-x_1)+3t_2(1+x_2)],\label{mn}\end{equation}
 and expressions for the effective masses
$m_{n\uparrow}, m_{n\downarrow}$ read (\cite{IY09})
\begin{eqnarray}
\frac{\hbar^2}{2m_{\uparrow(\downarrow)}}&=&\frac{\hbar^2}{2m_0}
+\frac{\varrho_{\uparrow(\downarrow)}}{2}
t_2(1+x_2)\label{m_ud}\\&\quad&+\frac{\varrho_{\downarrow(\uparrow)}}
{4}[t_1(1-x_1)+t_2(1+x_2)].\nonumber
\end{eqnarray}
In Eqs.~\p{mn}, \p{m_ud}, $t_i$ and $x_i$ are some phenomenological
parameters, specifying a given parametrization of the Skyrme
interaction. According to Eq.~\p{rho}, the number densities of
neutrons with spin up and spin down are determined by the spin
polarization parameter~$\Pi$, which, after the self-consistent
determination, is a function of the thermodynamic parameters
$\varrho,T,H$. Hence, the condition~\p{lowtemD} determines the
region in the domain of admissible values of the thermodynamic
parameters, where the entropy of a spin polarized state demonstrates
the expected behavior.

\begin{figure}[t]
\epsfxsize=\linewidth \epsfbox{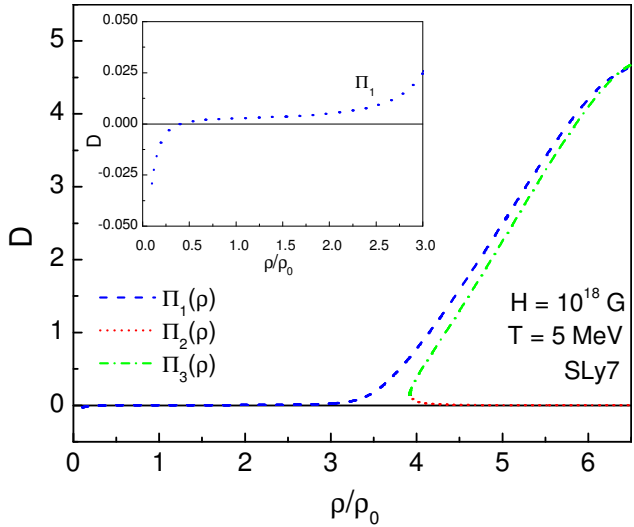} \caption{(Color online) The
difference $D$ in constraint~\p{lowtemD} for various branches
$\Pi_1$--$\Pi_3$ of spin polarization  as a function of density at
$H=10^{18}$~G and $T=5$~MeV for the SLy7 interaction.}\label{fig10}
\end{figure}

Fig.~\ref{fig10} shows the left-hand side $D$ of
constraint~\p{lowtemD} for the branches $\Pi_1$--$\Pi_3$ of spin
polarization as a function of density at $H=10^{18}$~G and
temperature $T=5$~MeV, at which the condition
$\varepsilon_{F\sigma}/T\gg 1$ holds true. For the branch $\Pi_1$,
the difference $D$ is negative up to the density
$\varrho_s\approx0.41\varrho_0$ (see the insert in
Fig.~\ref{fig10}), where it changes sign and remains positive for
all larger densities. Besides, for the branches $\Pi_2$ and $\Pi_3$,
the difference $D$ is positive for all densities at which the
corresponding solutions of the self-consistent equations exist.
These features explain the above mentioned unusual behavior of the
entropy for various  branches of spin polarization in neutron matter
with the Skyrme interaction under the presence of a strong magnetic
field. Note that such an unexpected behavior of the entropy was also
found for the states with spontaneous spin polarization (in the
absence of the magnetic field) in neutron matter  with the Skyrme
interaction~(\cite{RPV}) and antiferromagnetically spin ordered
states  in symmetric nuclear matter with the Gogny D1S
interaction~(\cite{IY04b,I05,I07}).

\section{CONCLUSIONS}

We have studied the impact of finite temperatures on the spin
structure in the magnetar interior, approximating the neutron star
matter by pure neutron matter and taking the Skyrme effective
interaction as a potential of NN interaction (SLy7
parametrization). According to the scalar virial theorem, strong
magnetic fields up to $10^{18}$~G can be relevant for the interior
regions of magnetars. It has been shown that, together with the
thermodynamically stable branch of solutions for the spin
polarization parameter corresponding to the case  when the
majority of neutron spins are oriented opposite to the direction
of the magnetic field (negative spin polarization), the
self-consistent equations, beginning from some threshold density,
have also two other branches  of solutions corresponding to
positive spin polarization. The influence of finite temperatures
on spin polarization remains moderate in the Skyrme model, at
least, up to  temperatures relevant for protoneutron stars. In
particular, a thermodynamic analysis, based on the calculation of
the free energy for different branches of spin polarization, shows
that the scenario with the metastable state characterized by
positive spin polarization, considered at zero
temperature~(\cite{IY09}), is preserved at finite temperatures as
well. This is one of the important conclusions of the given
research. The possible existence of a metastable state with
positive spin polarization will affect the neutrino opacities of a
neutron star matter in a strong magnetic field, and, hence, will
influence the cooling history of a neutron star~(\cite{RPLP}).

We have  also shown that above certain density
 the entropy for
various branches of spin polarization in neutron matter with the
Skyrme interaction in a strong magnetic field demonstrates the
unusual behavior being larger than that of the nonpolarized state.
To clarify this point, we have provided the corresponding
low-temperature analysis. It has been shown that this unexpected
behavior should be addressed to the dependence of the entropy of a
spin polarized state on  the effective masses of spin-up and
spin-down neutrons and to a certain inequality constraint on them
which is violated in the respective density range.

It is worthy to note that in the given research a neutron star
matter was approximated by pure neutron matter. This should be
considered as a first step towards a more realistic description of
neutron stars taking into account a finite fraction of protons
with the charge neutrality and beta equilibrium conditions.  In
particular, some admixture of protons can affect the onset
densities of enhanced polarization in a neutron star matter with
the Skyrme interaction~(\cite{IY04a}). Nevertheless, at such
strong magnetic fields, one can expect that the proton fraction is
relatively small and even can completely disappear in the dense
interior of a magnetar~(\cite{MKI}).

%--------------------------------------------------------------------
\acknowledgments

J.Y. was supported by grant 2010-0011378 from Basic Science
Research Program through NRF of Korea funded by MEST and by  grant
R32-2009-000-10130-0 from WCU project of MEST and NRF through Ewha
Womans University.
%--------------------------------------------------------------------

%-------------------------------------------------------------------
\end{document}